\documentclass[%
twocolumn, reprint,
 amsmath,amssymb,
 aps,
pra,
]{revtex4} 

\usepackage{graphicx}
\usepackage{dcolumn}
\usepackage{bm}
\usepackage{verbatim}
\usepackage[caption=false]{subfig}
\usepackage{CJK}
\usepackage{appendix}
\usepackage{color}

\begin{document}
\begin{CJK*}{UTF8}{}
\preprint{APS/123-QED}

\title{The dynamics of Tonks-Girardeau gas excited by a pulse drive}

\author{Jia Li%
\CJKfamily{gbsn}(李佳)}
\affiliation{Department of Physics and Institute of Theoretical Physics, University of Science and Technology Beijing, Beijing 100083, P.~R.~China}
\author{Yajiang Hao%
\CJKfamily{gbsn}(郝亚江)}
\email{haoyj@ustb.edu.cn}
\affiliation{Department of Physics and Institute of Theoretical Physics, University of Science and Technology Beijing, Beijing 100083, P.~R.~China}
%
\date{\today}

\begin{abstract}
In this paper we study the dynamics of Tonks-Girardeau (TG) gases in a harmonic potential driven by Gaussian pulse, which is a correspondence of the excitation dynamics of electrons in matters driven by ultrashort laser pulse. The evolving dynamics of TG gas are obtained with Bose-Fermi mapping method combined with the numerical techniques. We calculate the evolving dynamics of occupation distribution of single-particle energy levels, density distribution and momentum distribution of the system. It is shown that the system arrived at a dynamically stable state at the end of driving. At high-frequency regime TG gases return back to ground state while at low-frequency regime the population inversion exhibits and all atoms occupy high levels.

\end{abstract}

\maketitle

\end{CJK*}
\section{\label{sec:level1}INTRODUCTION}

Since cold atoms were realized in experiments, ultra-cold atoms have become a popular platform to investigate basic physical processes in many fields including condensed matter physics, quantum information processing, plasma physics, etc. Particularly, the rapid progress in cold atoms techniques make it play crucial roles to simulate many body quantum system. The experimental development of optical lattices deepens the understanding of not only the ground state quantum phase transition but also the non-equilibrium dynamics in periodical interacting system  \cite{EckardtRMP}. Cold atoms in periodically driven optical lattices were investigated as the important practical platform for Floquet engineering. The studies on the artificial gauge fields \cite{GaugeField}, the dynamical localization \cite{DynLoc, DynLoc2}, the dynamical quantum phase transition \cite{DynPhaTran}, topology physics \cite{TopBand, TopBand2, KWitersperger} have made great progress with the rapid advance in experiment techniques in cold atoms. Besides the dynamics in optical lattice \cite{Eck2017, Sor2005, Lig2007, Eck2009, Zen2009, Jot2014, Aid2015, Win2020, Col2019, Col2020, Oka2019, Gor2018,Kau2024, Rig2005}, modulated dynamics in harmonic trap \cite{EQuinn}, the temperature effect \cite{YYAtas}, the disorder effect for strong periodic forced bose gas \cite{Mar2024} and the nonequilibrium quantum thermodynamics \cite{YYAtas} were also studied. 

In cold atoms the atomic interaction can be tuned in the whole interacting regime from the weak interaction to the strong interaction with the Feschbach resonance and confined induced resonance techniques \cite{FR, CIR}. With anisotropic trap or optical lattices the cold atoms can be confined strongly in transverse direction and therefore the system becomes an one dimensional (1D) quantum system in the longitudinal direction \cite{Paredes, Toshiya, Ketterle, Single1D}. The tunability of atomic interaction, the controllability of dimension and time-dependent parameters offer us new opportunities that the traditional research areas cannot arrive at.  The 1D cold atoms system has been paid great attentions by both experimentalists and theorists \cite{RMP2011, RMP2012, RMP2013, Min2005}.

For Bose-Einstein condensates (BECs) with weak interaction we can obtain the ground state and dynamics by solving the Gross-Pitaevskii equation \cite{Dal1995}. The tunneling dynamics of BECs in a double-well can be described by two coupled nonlinear equations \cite{Sme1997}. When the atomic interaction is strong the system that can be solved analytically is very rare and the asymptotic analytical results exist only in specific limits. For example, Scopa et al. \cite{Sco2023} proposed an exact analytical formula for the single particle density matrix of the non-equilibrium Tonks-Girardeau (TG) gases \cite{Ton1936, Gir1960} at zero temperature in the limit of large $N$. Moreno et al. \cite{Mor2009} showed that the momentum distribution of TG gases at the high momentum limit is a power law form of $k^{-4}$, and their formula can replace numerical simulation in the high momentum region. For the 1D quantum system, non-perturbation method is necessary for its strong correlation effect. The generalized hydrodynamics method has been developed for the integrable system \cite{Sch2019, Ber2016, Cas2016} and is applicable in 1D cold atoms. Another effective approach for 1D system is the shortcuts to adiabaticity approach \cite{Has2024,Ode2023}.

Although the above experimental techniques and theoretical methods have have greatly stimulated the development of cold atom research, the new experiment protocols and theoretical tools are still expected. One of the important topics in the study on the quantum statistical physics, for example the eigenstates thermalization, is the preparation of long-lived excited states \cite{sTG,WKao}. By quenching the interaction across a confinement-induced resonance, the highly excited super-TG gases can be realized \cite{sTG}. Further, the stable excited state away from resonance regime can be realized by adding a weak dipolar repulsion among the atoms \cite{WKao}. So far, the dynamics of quantum gas are usually excited by quenching the atomic interaction or quenching the potential trap \cite{Collura2013,Joshua2020,Neel2021}. Le etc. apply a Bragg scattering pulse to observe the hydrodynamization and local prethermalization of 1D Bose gases \cite{Le2023}. In the present paper, we will study the evolving dynamics of TG gases in a harmonic potential driven by a Gaussian pulse, which is a correspondence of the excitation dynamics of electrons in matters driven by ultrashort laser pulse. It will be shown that under the pulse drive the TG gases will arrived at a dynamically stable excited state and occupation distribution inversion appears at the end of the pulse. The pulse optical drive is expected to be a candidate tool to obtain the stable excited state of TG gases.

In section \uppercase\expandafter{\romannumeral2}, we develop the numerical techniques based on the Bose-Fermi mapping method. In Section \uppercase\expandafter{\romannumeral3}, we show the explicit formula of the density distribution in coordinate space, the reduced one-body density matrix, the momentum distribution in momentum space and the occupation distribution of single particle levels. The numerical result for them are displayed in section \uppercase\expandafter{\romannumeral4}. Finally, the conclusion are given in Sec. V.

\section{Model and method}

We consider the dynamics of $N$ indistinguishable Bose atoms of mass $m$ in 1D that satisfy the following time-dependent Schr\"{o}dinger equation \cite{Gri2010}
\begin{equation*}
\begin{aligned}
    {\rm{i}}\hbar \frac{{\partial {\Psi _B}}}{{\partial t}} = &\sum\limits_{i = 1}^N {\left[ { - \frac{{{\hbar ^2}}}{{2m}}\frac{{{\partial ^2}}}{{\partial x_i^2}} + V\left( {{x_i},t} \right)} \right]} {\Psi _B} \\
    &+ {g_{1D}}\sum\limits_{1 \le i < j \le N} {\delta \left( {{x_i} - {x_j}} \right)} {\Psi _B},
\end{aligned}
\end{equation*}
in which the external potential is a harmonic potential superimposed by a drive potential, i.e.,
\begin{equation*}
    V\left( {x,t} \right) =\frac{1}{2} m\omega _0^2{x^2} - xS(t)\theta(t),
\end{equation*}
where $g_{1D}$ and ${\omega _0}$ are the effective 1D interacting constants and trapping frequency, respectively. The second term $xS(t)$ in $V(x,t)$ is a time-dependent drive, which is added to the system at time $t = 0$. $\theta(t)$ is unit step function, which is 0 and 1 for $t<0$ and $t \geq 0$, respectively. In the present paper, we will focus on the strongly interacting limit ${g_{1D}} \to +\infty $. The solution to this system can be obtained with the Fermi-Bose mapping method from the wavefunction of noninteracting fermions $\Psi_F(x_1,x_2,\cdots,x_N;t)$ \cite{Bul2007}
\begin{equation*}
    {\Psi _B}\left( {{x_1}, \cdots ,{x_N};t} \right) =A\left( {{x_1}, \cdots ,{x_N}} \right) {\Psi _F}\left( {{x_1}, \cdots ,{x_N};t} \right)
\end{equation*}
with the mapping function
\begin{equation*}
    A\left( {{x_1}, \cdots ,{x_N}} \right) = \prod\limits_{1 \le i < j \le N} {{\rm{sign}}\left( {{x_i} - {x_j}} \right)}.
\end{equation*}
Its effect is to map the exchange antisymmetric wave function of fermions into an exchange symmetric wave function of bosons. sign($x$) is sign function and is equal to 1, 0, and -1, for $x$ $>0$, $=0$, and $<0$, respectively. We just need to solve the Schr\"{o}dinger equation,
\begin{equation}
    {\rm{i}}\hbar \frac{{\partial {\Psi _F}}}{{\partial t}} = \sum\limits_{i = 1}^N {\left[ { - \frac{{{\hbar ^2}}}{{2m}}\frac{{{\partial ^2}}}{{\partial x_i^2}} + \frac{1}{2}m\omega _0^2{x_i^2} - {x_i}S(t)} \right]} {\Psi _F}
\label{eq2}
\end{equation}
with respect to ${\Psi _F}\left( {{x_1}, \cdots ,{x_N};t} \right)$, and ${\Psi _B}\left( {{x_1}, \cdots ,{x_N};t} \right)$ is obtained by Bose-Fermi mapping.

The general wave function to the above time-dependent problem can be formulated as
\begin{equation}
    {\Psi _F} = \sum_n {{c_n}{\psi _n}\left( {{x_1}, \cdots ,{x_N}} \right){e^{ - {\rm{i}}{E_n}t/\hbar }}},
\label{eq1}
\end{equation}
where $n$ specify different eigen functions ${\psi _n\left( {{x_1}, \cdots ,{x_N}} \right)}$ with ${E_n}$ being the corresponding eigen energy and ${c_n}$ is the superposition coefficient. The energies ${E_n}$ are listed in order from the smallest to the largest. Theoretically it is a summation of infinite terms, but it is reasonable to assume that in a finite time the probability is very small for a particle to be excited to very high energy levels. So we can truncate the infinite eigen space into a finite space. The eigen function is compactly written in a form of the Slater determinant,
\begin{equation*}
    {\psi _n}\left( {{x_1}, \cdots ,{x_N}} \right) = \frac{1}{{\sqrt {N!} }}\mathop {\det }\limits_{l,k = 1}^N \left[ {{\varphi _{l - 1}}\left( {{x_k}} \right)} \right],
\end{equation*}
where ${\varphi _l}(x)$ is the 1D harmonic oscillator eigenfunction. The different eigen functions  correspond to different atom occupation configuration and can be simply denoted by the Dirac symbol $\left|\psi _1\right\rangle = \left| {0,1 \cdots, N - 2, N - 1} \right\rangle$, $\left|{\psi _2}\right\rangle = \left| {0,1, \cdots N - 2,N} \right\rangle$, $\left|{\psi _3}\right\rangle = \left| {0,1, \cdots N - 2,N + 1} \right\rangle$, $\left|{\psi _4}\right\rangle = \left| {0,1, \cdots N - 3,N - 1,N} \right\rangle$, and etc.. Here the set of numbers in $\left|n_1,n_2,\cdots,n_N\right\rangle$ denote the occupied single particle energy levels of harmonic trap.

Combining Eq. (\ref{eq1}) with Eq. (\ref{eq2}), we have
\begin{equation*}
    {\rm{i}}\hbar {{\dot c}_m} =  - \sum\limits_{n = 1} {{c_n}(t){\Gamma _{mn}}} {e^{ - {\rm{i(}}{E_n} - {E_m})t/\hbar }}
\end{equation*}
with the transition matrix elements
\begin{equation}
    {\Gamma _{mn}} = S(t)\sum\limits_{j = 1}^N {\int {{\psi _m}{x_j}{\psi _n}} } \prod\limits_{i = 1}^N {{\rm{d}}{x_i}}.
\label{eqa1}
\end{equation}
It can be proved that $\Gamma $ is a sparse matrix (Appendix A)
\begin{eqnarray*}
&&{\Gamma _{mn}} =          \\
&&\begin{aligned}
    & \frac{{{\alpha _0}}}{{\sqrt 2 }}S(t)\sum\limits_{i = 1}^{N} {\sqrt {{n_i} + 1} \left| {\begin{array}{*{20}{c}}
    {{\delta _{{n_1},{m_1}}}}&{{\delta _{{n_i} + 1,{m_1}}}}&{{\delta _{{n_{N}},{m_1}}}}\\
    {{\delta _{{n_1},{m_i}}}}&{{\delta _{{n_i} + 1,{m_i}}}}&{{\delta _{{n_{N}},{m_i}}}}\\
    {{\delta _{{n_1},{m_{N}}}}}&{{\delta _{{n_i} + 1,{m_{N}}}}}&{{\delta _{{n_{N}},{m_{N}}}}}
    \end{array}} \right|} \\
    +& \frac{{{\alpha _0}}}{{\sqrt 2 }}S(t)\sum\limits_{i = 1}^{N} {\sqrt {{n_i} } \left| {\begin{array}{*{20}{c}}
    {{\delta _{{n_1},{m_1}}}}&{{\delta _{{n_i} - 1,{m_1}}}}&{{\delta _{{n_{N}},{m_1}}}}\\
    {{\delta _{{n_1},{m_i}}}}&{{\delta _{{n_i} - 1,{m_i}}}}&{{\delta _{{n_{N}},{m_i}}}}\\
    {{\delta _{{n_1},{m_{N}}}}}&{{\delta _{{n_i} - 1,{m_{N}}}}}&{{\delta _{{n_{N}},{m_{N}}}}}
    \end{array}} \right|},
\end{aligned}
\end{eqnarray*}
with ${\alpha _0} = \sqrt {{\hbar  \mathord{\left/{\vphantom {\hbar  {m{\omega _0}}}} \right.\kern-\nulldelimiterspace} {m{\omega _0}}}}$. It can be seen from the equations of the transition matrix element that the transition process of multiparticles is highly similar to that of single particle. The selection rule of the harmonic oscillator suggests that a particle can only transit to its nearest neighboring energy level, so dipole transition are not allowed. This rule is expanded in the multi-particle transition process, which is fundamentally due to the existence of a high degenerate of simplicity of multi-particle energy levels in a simple harmonic potential well, leading to an increase in the range of neighboring energy levels. For simplicity, in the following numerical calculations we will take $m = \hbar  = {\omega _0} = {\alpha _0} = 1$ and the dimension of time is $1/\omega _0$.

The present numerical approach based on Bose-Fermi mapping method is applicable for TG gases drived by arbitrary external potential besides the periodical driving \cite{Hao2023}.

\section{The density profile and momentum distribution}

The density distribution of many particles in real space is defined as
\begin{eqnarray}
    \rho (x,t) &&= N\int {\rm{d}}x_2{\rm{d}}x_3\cdots {\rm{d}}x_N  \label{eq4} \\
    &&\times \Psi _B^*(x,x_2, \cdots ,x_N;t)\Psi _B(x,x_2, \cdots ,x_N;t).      \nonumber
\end{eqnarray}
Following the definition in Sec. II, the density distribution can be formulated as
\begin{equation*}
\begin{aligned}
    \rho (x,t) = N\sum\limits_{m = 1}^N {\sum\limits_{n = 1}^N {c_m^*} } {c_n}{D_{mn}}{e^{{\rm{i}}\left( {{E_m} - {E_n}} \right)t}},
\end{aligned}
\end{equation*}
in which (see Appendix B for details)
\begin{equation*}
\begin{aligned}
    {D_{mn}} = \frac{1}{N}\sum\limits_{i = 1}^N {\sum\limits_{j = 1}^N {{\varphi _{{n_i}}}{\varphi _{{m_j}}}} \left( x \right){{\left( { - 1} \right)}^{i + j}}\prod\limits_{k,l \ne i,j}^N {{\delta _{{n_k},{m_l}}}} }.
\end{aligned}
\end{equation*}

With the many-body wave function, the reduced onebody density matrix (ROBDM) is formulated as
\begin{equation}
\begin{aligned}
    \rho (x,y,t) &= N\int {\rm{d}}x_2{\rm{d}}x_3\cdots {\rm{d}}x_N \\
    &\times\Psi _B^*(x, \cdots ,{x_N};t)\Psi _B(y, \cdots ,{x_N};t)
\label{eq8}
\end{aligned}
\end{equation}
and following the same procedure as the above calculation of density distribution $\rho (x,y,t)$ can be expressed as
\begin{equation*}
    \rho (x,y,t) = N\sum\limits_{m = 1}^N {\sum\limits_{n = 1}^N {c_m^*} } {c_n}{P_{mn}}{e^{{\rm{i}}\left( {{E_m} - {E_n}} \right)t}}.
\end{equation*}
Here $P_{mn}$ is the integral of different eigen functions
\begin{equation*}
\begin{aligned}
    {P_{mn}} =& \int \prod\limits_{i = 2}^N {{\rm{d}}{x_i}}  {{\rm{sign}}\left( {x - {x_i}} \right){\rm{sign}}\left( {y - {x_i}} \right)} \\
    &\times {{\psi _m}\left( {x, \cdots ,{x_N}} \right){\psi _n}\left( {y, \cdots ,{x_N}} \right)}
\end{aligned}
\end{equation*}
and the integral can be given in a concise form (see Appendix B for details) \cite{Pez2007}
\begin{equation*}
    {P_{mn}} = \sum\limits_{i = 1}^N {\sum\limits_{j = 1}^N {{\varphi _{{n_i}}}(x)} {\varphi _{{m_j}}}(y){A_{ij}}}
\end{equation*}
with
\begin{equation*}
    {A_{ij}} = {( - 1)^{i + j}}\det {{{M}}_{kl}}
\end{equation*}
and
\begin{equation*}
    {M_{kl}}(x,y) = {\delta _{kl}} - 2\int_x^y {{\varphi _k}(\alpha ){\varphi _l}(\alpha ){\rm{d}}\alpha }.
\end{equation*}

While we obtain the ROBDM, the momentum distribution can be obtained by its Fourier transformation
\begin{equation*}
    n(k,t) = \frac{1}{{2\pi }}\int {{\rm{d}}x} \int {{\rm{d}}y\rho (x,y,t){e^{ - {\rm{i}}k(x - y)}}}.
\end{equation*}

It is also interesting to investigate the occupation distribution \cite{For2003} evolution of the single particle eigen states of harmonic potential induced by the external driving, which is denoted by $p_j$ in the following. Obviously, summing the number of particles over all energy levels give the total number of particles $N$. The relationship between $p_j$ and $c_n$ is
\begin{equation*}
    {p_j} = \sum\limits_{n} {{{\left| {{c_n}} \right|}^2}}p_{nj},
\end{equation*}
if the energy level $j$ is a harmonic oscillator state of the constituent quantum states $\left|{\psi _n}\right\rangle$ then $p_{nj}=1$, otherwise $p_{nj}=0$.

\section{Numerical results}

\begin{figure}[htbp]
    \subfloat{\includegraphics[width=8cm,height=3.450cm]{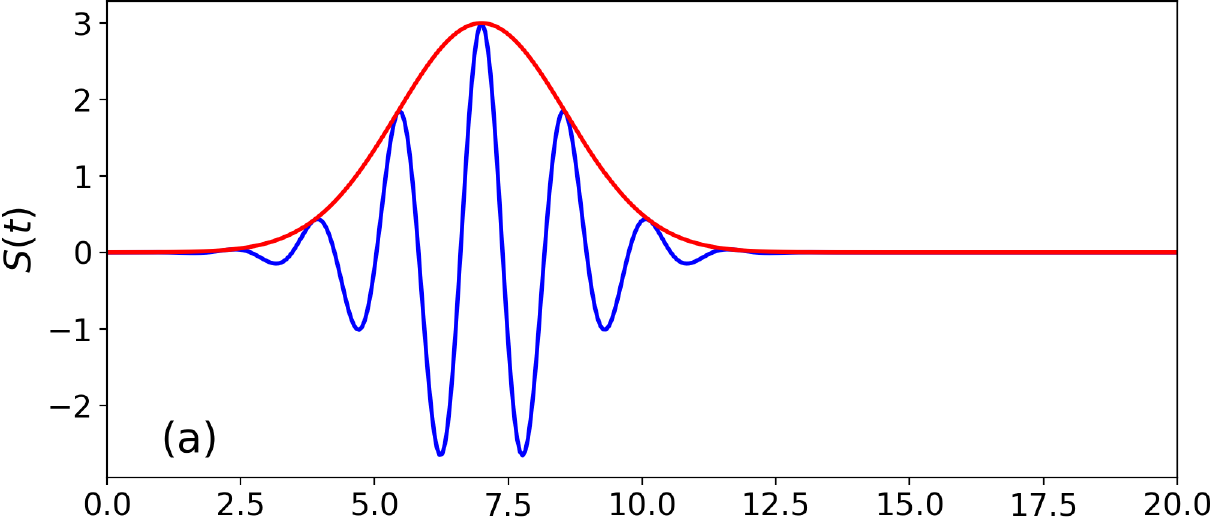}}
    \vspace{0in}
    \subfloat{\includegraphics[width=8cm,height=2.201cm]{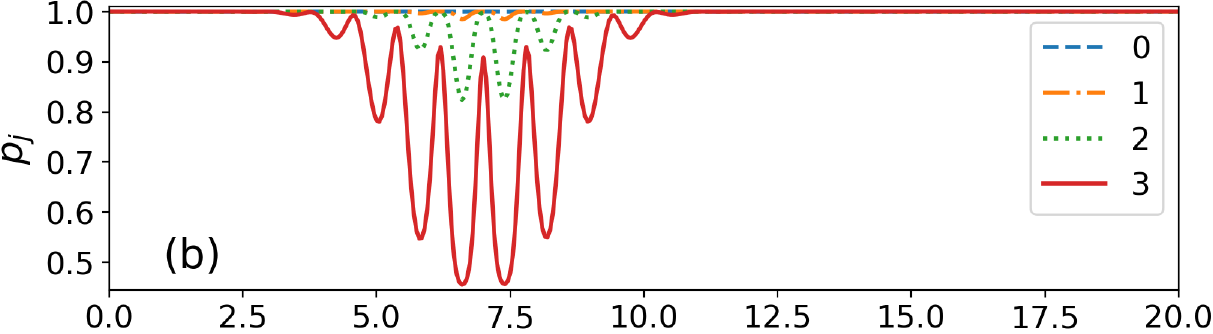}}
    \vspace{0in}
    \subfloat{\includegraphics[width=8cm,height=2.201cm]{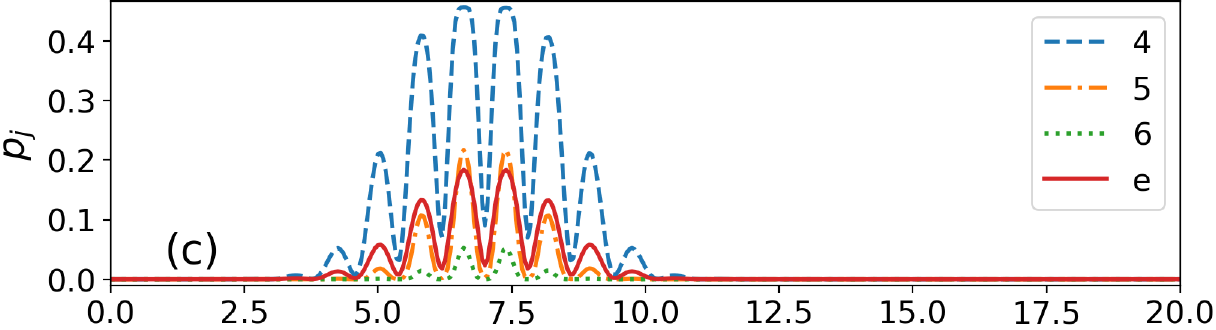}}
    \vspace{0in}
    \subfloat{\includegraphics[width=8cm,height=3.463cm]{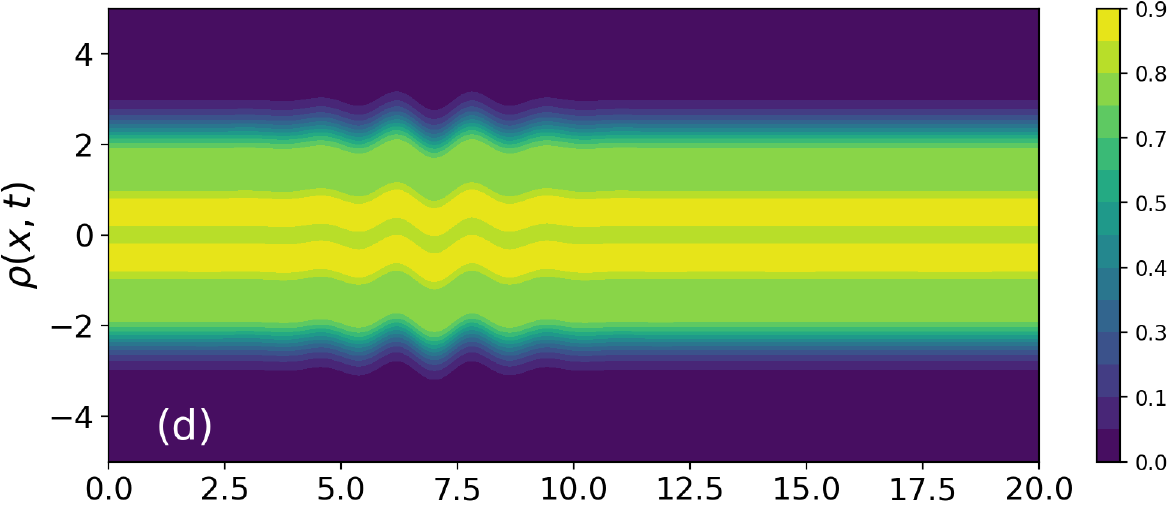}}
    \vspace{0in}
    \subfloat{\includegraphics[width=8cm,height=3.822cm]{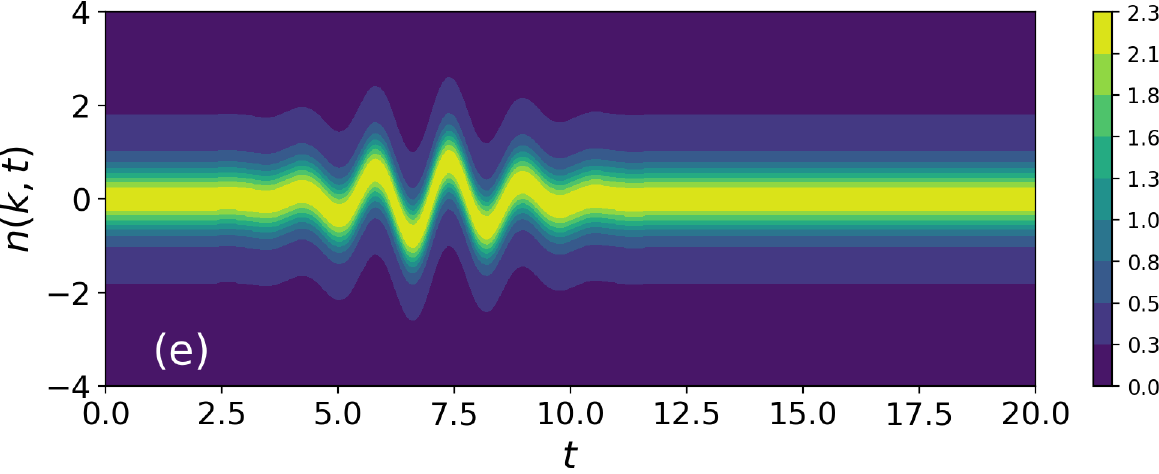}}
    \caption{The evolving dynamics of TG gases with $N$ = 4, $S_A$=3, $\Delta$ = 5 and $\omega$ = 4 $\omega _0$. (a) The laser pulse $S(t)$; (b) The occupations of the initially occupied levels: $p_j$ for $j$ = 0, 1, 2, and 3; (c) The occupations of the initially unoccupied levels: $p_j$ for $j$ = 4, 5, 6 and the total occupations $p_e$; (d) The density distribution $\rho(x,t)$; (e) The momentum distribution $n(k,t)$.}
    \label{fig1}
\end{figure}

In this section, we will show the evolution of density distribution, momentum distribution and occupation distribution of single particle energy levels induced by the external drive potential. In order to solve the problem of high computational cost because of the enormous eigen space, we truncate the eigen space at a certain energy threshold (the total energy of the specified eigen state of many particles), and thus greatly reduce the dimension of Hilbert space.

In the present paper we focus on a non-periodical driving to simulate the excited dynamics of matter radiated by ultrashort pulse laser \cite{Dup2023, Dup2021}. For a general drive pulse $S(t) = S_A\exp \left( { - {{(t - t_0)}^2}/\Delta} \right)\sin \left( {\omega(t - t_0) + \phi _0} \right)$, where $S_A$ is the driving strength, $\omega /2\pi$ is the center or carrier frequency and $\sqrt{\Delta /2}$ determines the bandwidth of the pulse, such as the drive shown in Fig. \ref{fig1}a, in which the blue solid lines display the pulse drive of a Gaussian envelope (the red solid lines). In the subsequent calculation, we stipulate that $t_0 =7$ and $\phi _0 = \pi /2$, which are not sensitive for the present study.

The occupations of single particle energy level, the density distributions, and momentum distributions for \textcolor{red}{TG} gases of $N = 4$ are displayed in Fig. \ref{fig1}b-Fig. \ref{fig1}d. Initially, the TG gases stay at its ground state in which four atoms occupy at the lowest four energy levels with $p_j=1$ ($j=0,1,2,3$). Under the external drive, the atom at the highest occupied energy level (HOEL) ($j=3$ here) is excited at first and the excitation of the atoms at lower energy levels follow up. The occupations of initially occupied levels decrease oscillatingly in the first half of Gaussian drive and increase oscillatingly in the second half. The occupation distribution evolution of the initial unoccupied levels is just the reverse. The lowest unoccupied energy level (LUEL) ($j=4$ here) is occupied firstly and those higher unoccupied levels are occupied successively. During the evolution the occupation of HOEL and LUEL exhibit the larger amplitudes comparing with other initial occupied energy levels and unoccupied energy levels. The total occupations of all single particle excited states ($p_e=\sum_{j\geq N} p_j/N$) are also ploted in red solid lines in Fig \ref{fig1}c (the total atom number is divided here to match the whole figure). During the pulse drive the oscillation amplitude of occupations increase and decrease as the pulse drive become strong and weak. All the evolution take place during the pulse drive and the evolution dynamics stop at the end of the pulse drive. It is interesting to notice that the whole system evolves back into the initial states completely once the pulse drive closes. 

The density distribution (Fig. \ref{fig1}d) and momentum distributions (Fig. \ref{fig1}e) also oscillate strongly or weakly with the change of the pulse drive amplitude. The center of the distribution oscillates with the change of pulse drive but both of them preserve their initial distribution properties. In the coordinate space the TG gases behave in the same way as noninteracting fermions and exhibit a shell structure of $N$ peaks. While in momentum space they display single peak structure that is a typical property of bosons \cite{Cha2024}.

\begin{figure}[htbp]
    \centering
    \subfloat{\includegraphics[width=8cm,height=2.201cm]{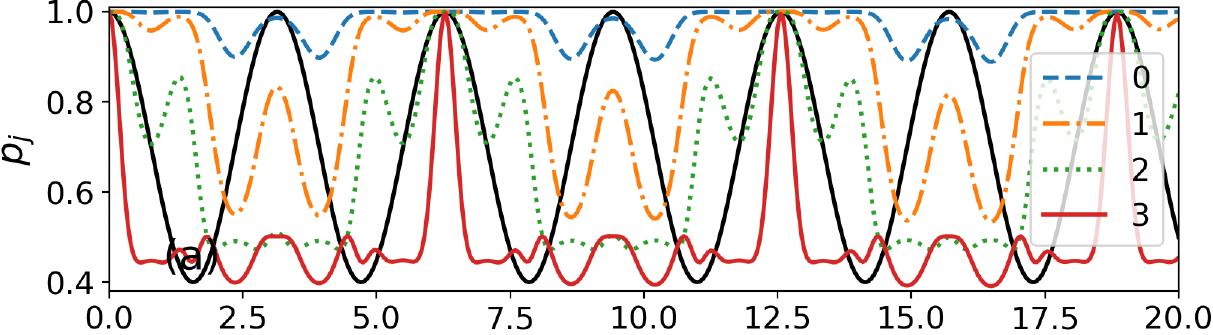}}
    \vspace{0in}
    \subfloat{\includegraphics[width=8cm,height=2.201cm]{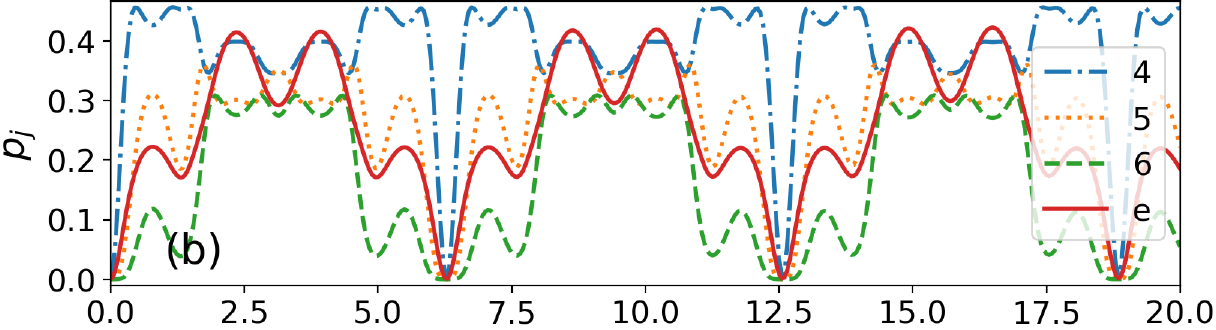}}
    \vspace{0in}
    \subfloat{\includegraphics[width=8cm,height=3.822cm]{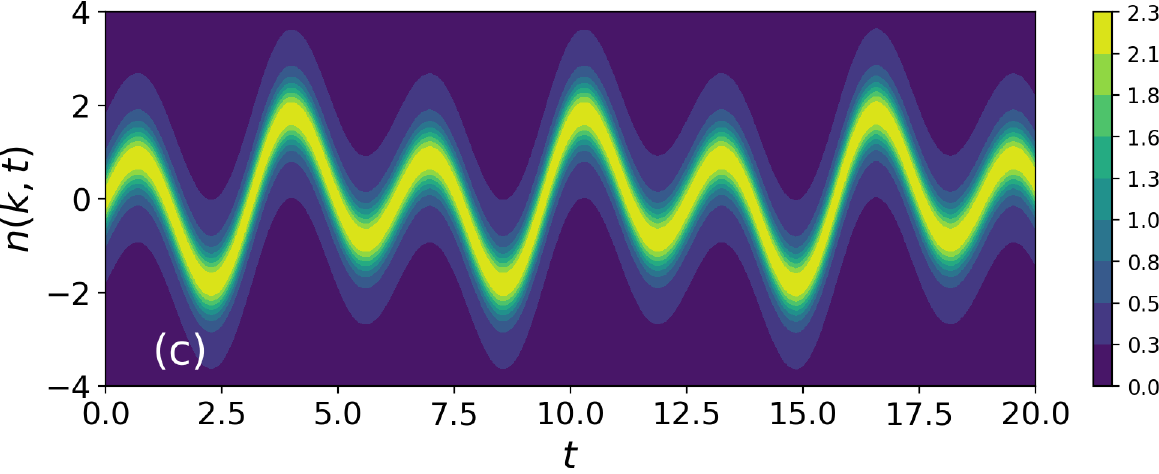}}
    \caption{The evolving dynamics of TG gases with $N$ = 4, $S_A$=2, $\Delta$ = $\infty$ and $\omega$ = 2$\omega _0$. (a) The occupations of the initially occupied levels: $p_j$ for $j$ = 0, 1, 2, and 3; (b) The occupations of the initially unoccupied levels: $p_j$ for $j$ = 4, 5, 6 and the total occupations $p_e$; (c) The momentum distribution $n(k,t)$. The solid black line in (a) shows the scaling of the drive $S(t)$ over time.}
    \label{fig2}
\end{figure}

In the following we investigate the evolution dynamics under the drive potential in two limiting case, i.e., the periodical drive $S(t) = S_A\sin \left( {\omega t + \phi _0} \right)$ (in the infinite bandwidth limit $\Delta \rightarrow \infty$) and the Gaussian drive (in the zero frequency limit $\omega \rightarrow 0$).

The density distribution in coordinate space will not be displayed because TG gases always oscillate as an entity with the same properties as those of the pulse drive. We plot the evolving dynamics of the occupations of single particle energy levels and momentum distribution for TG gases driven by the periodical potential in Fig. \ref{fig2}. It is shown that all of them evolve periodically with the period being twice that of the drive potential. It is same as the case in Fig. \ref{fig1}, in which the occupation of HOEL oscillate at first with a larger amplitude and that of other lower initially occupied level start to oscillate later with smaller amplitude. For the initial occupied levels, the higher the level, the greater the amplitude and the earlier the oscillation begins. For the initially unoccupied levels, the lower the level, the greater the amplitude and the earlier the oscillation begins. The occupation of LUEL start to oscillate earliest and has the maximum amplitude. The momentum distribution evolve periodically with the same period as that of occupation. During the evolution the momentum distribution preserve the typical single peak structure of bosons. The only difference from those in Fig. \ref{fig1} is that the dynamics are periodical in the present situation. The above results are consistent with those obtained by the analytical method \cite{Hao2023}, which fully demonstrates the correctness of the present numerical method and it can be calculated for arbitrary external drive.

\begin{figure}[htbp]
    \centering
    \subfloat{\includegraphics[width=8cm,height=2.201cm]{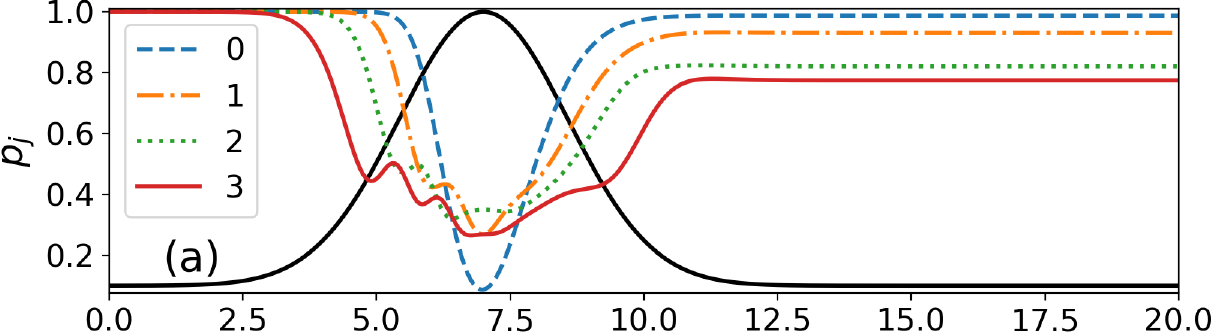}}
    \vspace{0in}
    \subfloat{\includegraphics[width=8cm,height=2.201cm]{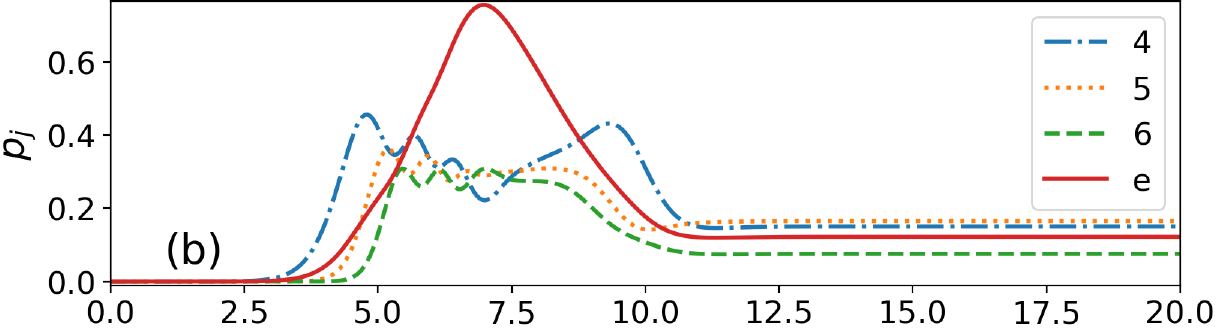}}
    \vspace{0in}
    \subfloat{\includegraphics[width=8cm,height=3.822cm]{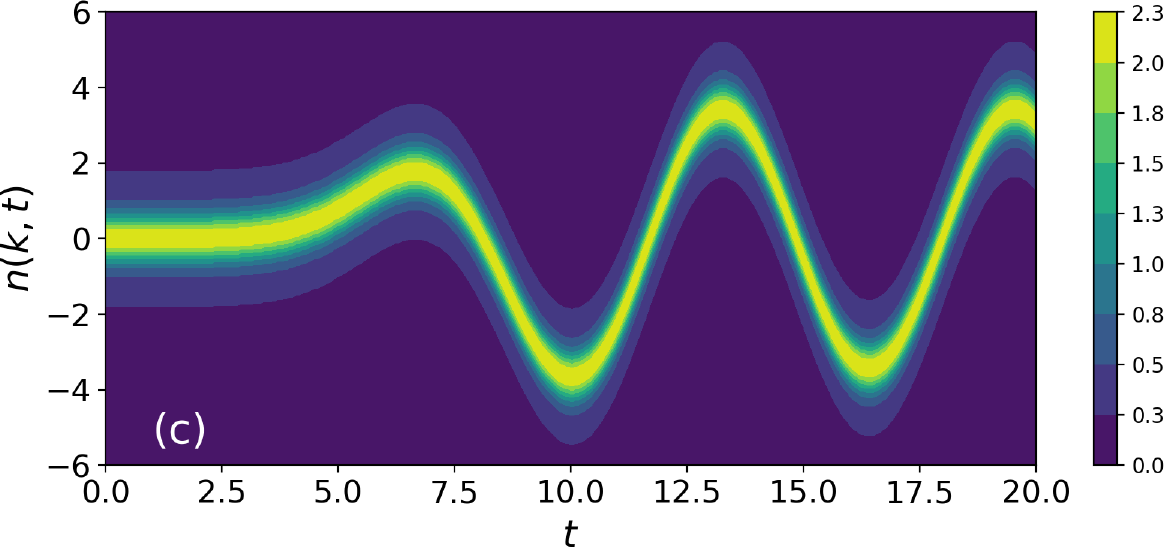}}
    \caption{The evolving dynamics of TG gases with $N$ = 4, $S_A$ = 3, $\Delta$ = 5 and $\omega$ = 0. (a) The occupations of the initially occupied levels: $p_j$ for $j$ = 0, 1, 2, and 3; (b) The occupations of the initially unoccupied levels: $p_j$ for $j$ = 4, 5, 6 and the total occupations $p_e$; (c) The momentum distribution $n(k,t)$. The solid black line in (a) shows the scaling of the drive $S(t)$ over time.}
    \label{fig3}
\end{figure}

When $\omega \rightarrow 0$, the pulse becomes a single Gaussian pulse drive $S(t) = S_A\exp \left( { - {{(t - t_0)}^2}/\Delta} \right)$. The evolution of occupations and momentum distribution are plotted in Fig. \ref{fig3} for Gaussian drive. In this situation, the evolution of occupation still begin at the HOEL and the LUEL. The occupations of initially occupied levels are shown in Fig. \ref{fig3}a. They decrease at the first half of the Gaussian drive and arrive at the smallest as the Gaussian drive approach to the strongest. Later the occupations increase and reach the stable values at the end of the pulse drive and the occupation of low level is higher than that of high level. The occupation evolution of the initially unoccupied level is plotted in Fig. \ref{fig3}b. These levels also arrive at a stable occupations at the end of drive. It is interesting that the LUEL is not occupied with the largest probability finally. The total occupation of all initial unoccupied levels is also exhibited in Fig. \ref{fig3}b in red solid lines, which arrive at the peak value near the maximum drive and become stable as the Gaussian drive terminates. In a word, the Gaussian pulse drive excite the TG gases to the excited state in which atoms populate in more energy levels with definite probabilities. In this case the momentum distribution still exhibits the typical single peak structure of bosons during the evolution, which is plotted in Fig. \ref{fig3}c. Because the TG gas is excited, its momentum distribution oscillates periodically around the zero momentum and arrives at a dynamical stable state. This is obviously different from the cases in Fig. \ref{fig1} and in Fig. \ref{fig2}.

\begin{figure}[htbp]
    \centering
    \subfloat{\includegraphics[width=8cm,height=3.653cm]{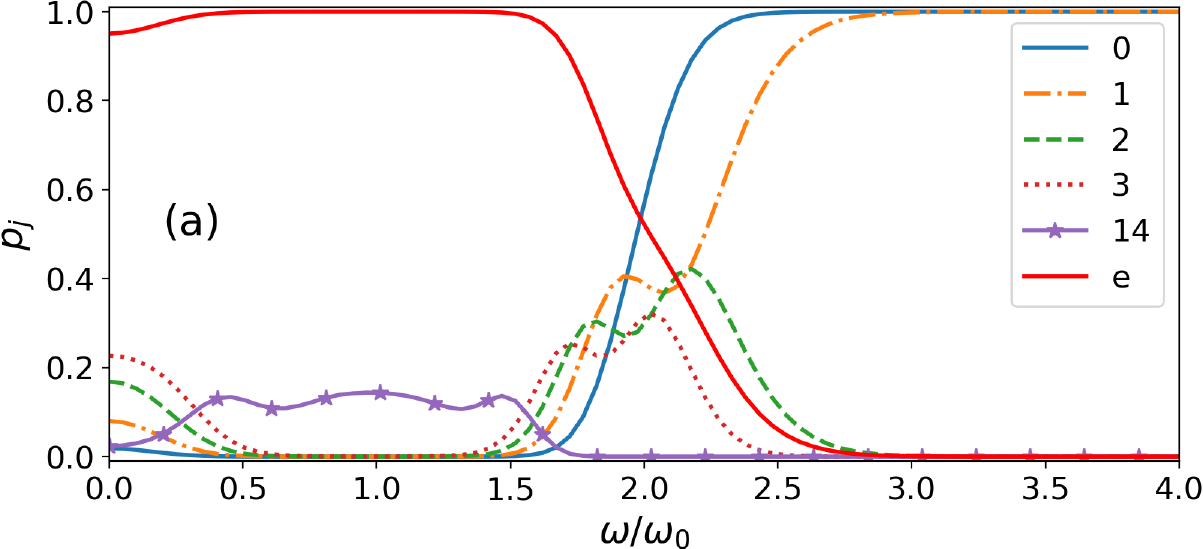}}
    \vspace{0in}
    \subfloat{\includegraphics[width=8cm,height=3.747cm]{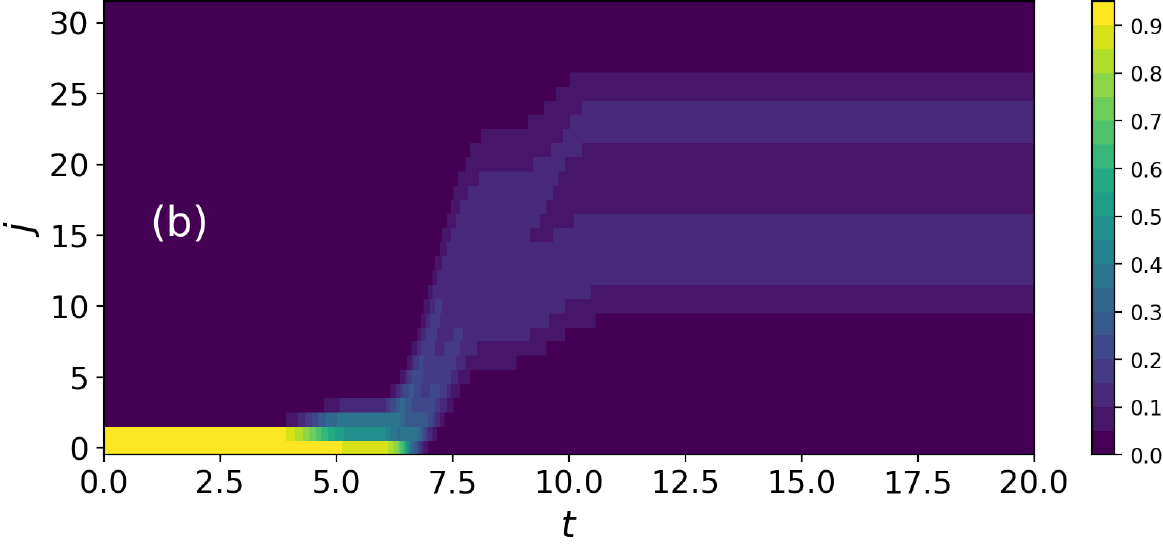}}
    \caption{The occupations for the TG gases with $N$ = 2, $S_A$=3 and $\Delta$ = 5. (a) The final stable occupations dependence on the pulse driving frequency; (b) The evolving occupation distribution for $\omega$ = $\omega_0$.}
    \label{fig4}
\end{figure}

The numerical calculations demonstrate that occupation distribution of energy levels always arrive at stable as long as the external drive come to the end. It is only under the periodical drive that the occupation distribution oscillates periodically. It is interesting to investigate the final stable occupation distribution of different levels after the end of pulse drive. We display the occupation distribution dependence on the driving frequency $\omega$ in Fig. \ref{fig4}a for $N=2$, in which the occupation of lower levels $p_j$ ($j=0,1,2,3$) and the total occupation of high levels ($j>2$) are plotted, and as a representative of high levels $p_{14}$ is also plotted. It is shown that under the high frequency driving ($\omega>3\omega_0$) the TG gases return back to the ground state and $N$ atoms occupy the lowest $N$ levels with one atom being on one level. Under the low frequency driving, the occupation inversion take place. In the region of $\omega<0.5\omega_0$, atoms are excited to high levels ($j>N-1$) with larger probability and occupy the low levels ($j=0,...,N-1$) with smaller probability. As the driving frequency is close to trap frequency of harmonic trap ($0.5\omega_0 <\omega < 1.5\omega_0$) all atoms occupy high levels (for example, the level $j=14$) and $p_e$ arrive at 1, while the lower levels are vacant ($j=0,1,2,3$). The exciting process of atoms from the low levels to high levels for $\omega=\omega_0$ is displayed in Fig. \ref{fig4}b. In the first period (around $t<6$) the occupation of HOEL decreases at first followed by the decrease of level $j=0$, and the occupation of HUEL increases at first followed by the increase of higher levels sequentially. In the second period (around $t=9$), atoms distribute on levels of $8\leq j\leq 19$ with almost equal probability. Finally the occupation evolve into two branches with upper levels around $j=24$ and down levels around $j=14$.

\begin{figure}[htbp]
    \centering
    \subfloat{\includegraphics[width=4cm,height=3.195cm]{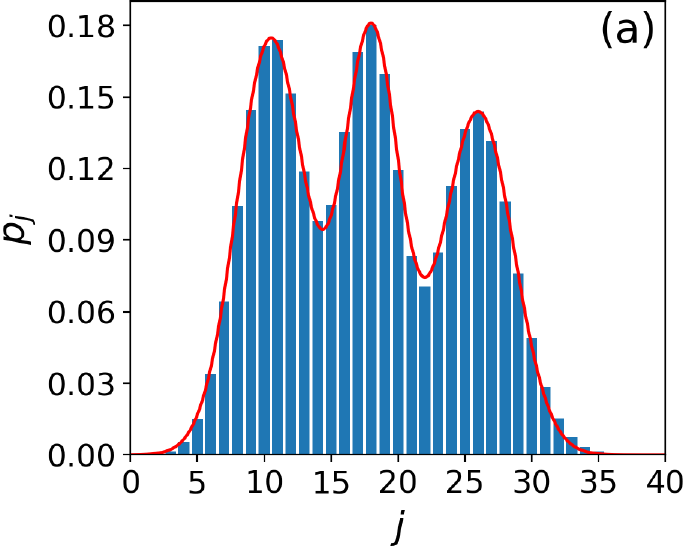}}
    \hspace{0in}
    \subfloat{\includegraphics[width=3.748cm,height=3.2cm]{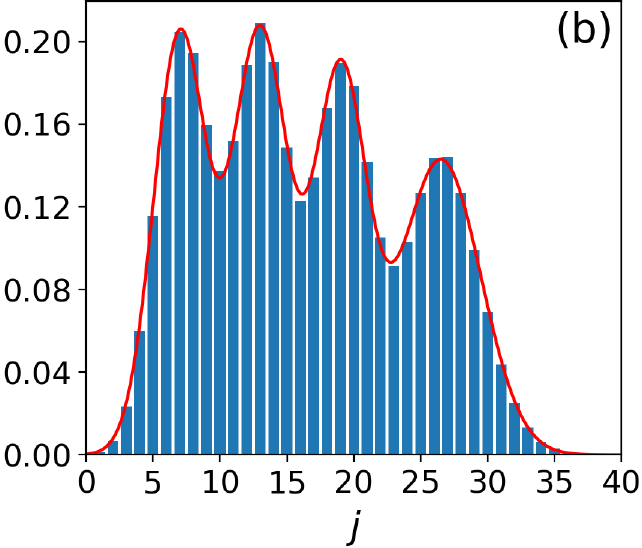}}
    \caption{The occupations at the end of the pulse when $\omega =\omega_0$ for different particle numbers: $N=3$ (a) and $N=4$ (b). The blue bar is the calculated data, and the red line is the fitting curve of Gaussian function.} 
\label{fig5}
\end{figure}

When the driving frequency is close to the level space of harmonic potential, i.e., $\omega  \sim \omega_0$, no particle occupies the lowest $N$ levels and all atoms are completely excited to high levels (complete excitation). The peak number of final stable multi-peak structure is the number of atoms. In Fig. 5 we display the occupation distribution of $N$=3 and 4. Each peak is a Gaussian, which can be confirmed by fitting the occupations to the Gaussian functions $p(j) = \sum\limits_{i = 1}^N {{a_i}\exp \left( { - {{{\left( {j - {b_i}} \right)}^2}}/{{c_i}}} \right)}$. The fitted lines are plotted in Fig. \ref{fig5} in red solid lines. The special occupation distribution is related with the larger transition coefficient for the higher energy level. The higher transition result in the faster transition, so that after a longer excitation, each particle displays a separate peak. 

\begin{figure}[htbp]
    \centering
    \subfloat{\includegraphics[width=8cm,height=3.637cm]{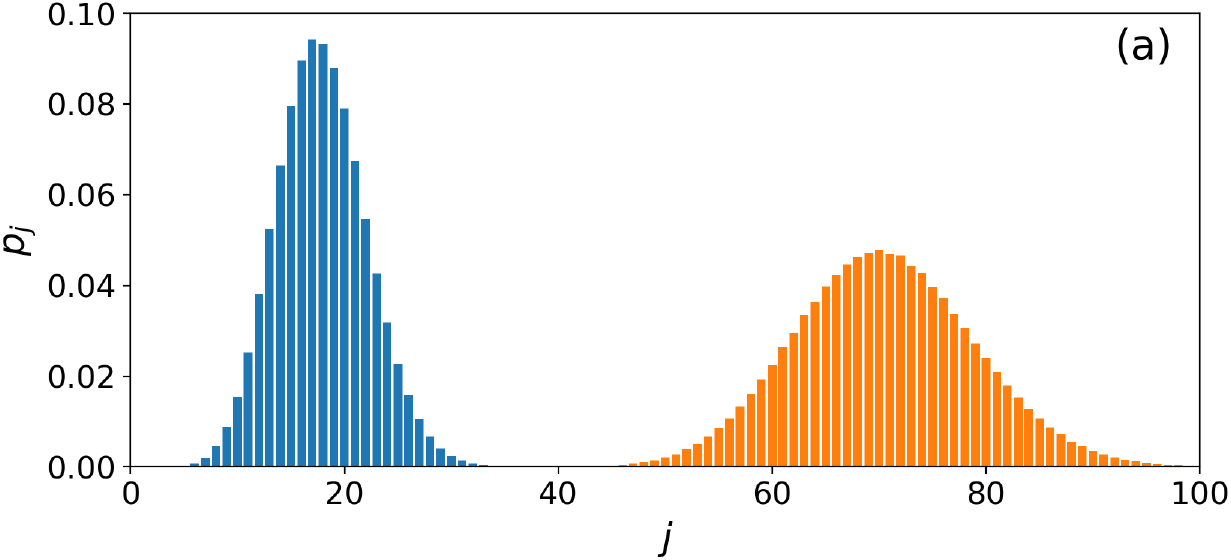}}
    \hspace{0in}
    \subfloat{\includegraphics[width=8cm,height=3.326cm]{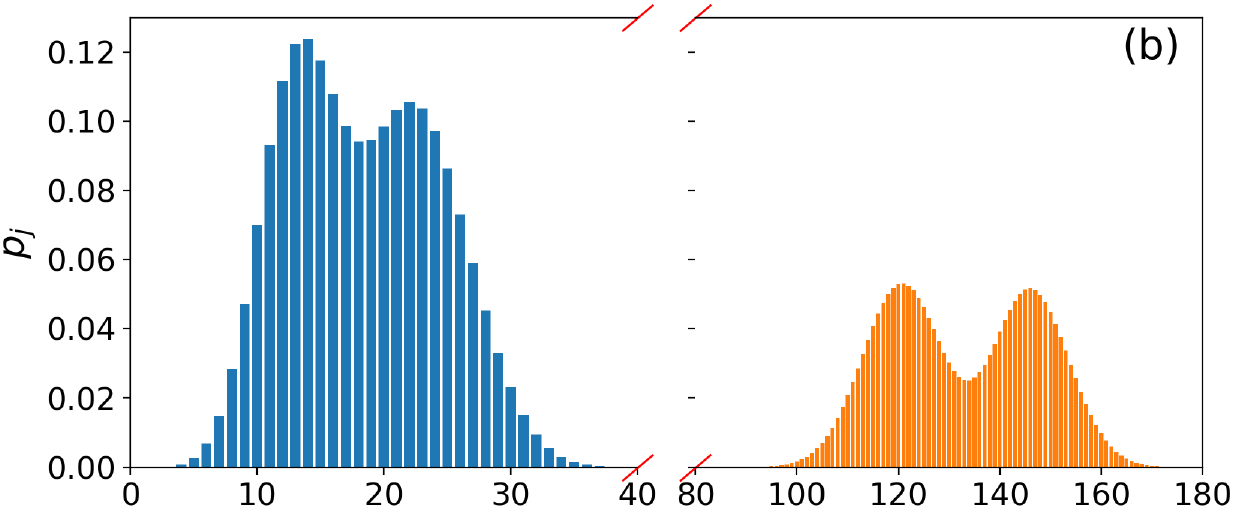}}
    \caption{The final occupation: The orange bar doubles the pulse intensity relative to the blue bar for $N=1$ (a); The orange bar doubles the pulse intensity and pulse width relative to the blue bar for $N=2$ (b).}
    \label{fig6}
\end{figure}

In complete excitation, the distribution of occupy number is no longer sensitive to the change of pulse parameters, and always presents a Gaussian distribution. However, the higher the pulse intensity and pulse width, the overall distribution will move to the higher energy level. At the same time, due to the continuous increase of the transition coefficient, the Gaussian distribution of the occupancy number will become shorter and wider (Fig. \ref{fig6}a), and the distance between the multi-particle peak centers will also increase (Fig. \ref{fig6}b).

\section{Conclusion}

We study the dynamics of TG gases confined in a harmonic potential driven by a pulse driving. The time-dependent many-body wave function is obtained based on the Bose-Fermi mapping method combined with numerical method. The occupation distribution of single particle levels, and the density distribution and momentum distribution are exhibited. It is shown that during driving the density distribution always exhibits a shell structure and momentum distribution exhibits a typical single peak structure and the structures will not deform. The TG gases oscillate around the initial position as a whole with the change of driving potential. The oscillation amplitude also increase and decrease with the increase and decrease of the driving strength except that under the periodical driving. The occupation evolution of single particle energy levels always begin from the HOEL and LUEL and both of them has the biggest oscillation amplitude. It is interesting to note that under pulse drive TG gases always arrive at a dynamically stable state in which the momentum distribution oscillates around the point $k=0$ periodically and the occupation distribution exhibit the behaviour of multi-Gaussian distribution with peak number being the atom number. Under the high-frequency driving TG gases return back to ground state 
at the end of the driving and under the low-frequency driving TG gases are excited to stable excited states and occupation inversion is displayed. As the driving frequency approximate to the trap frequency atoms are excited completely and the lower levels will not be occupied. 
In addition, the method used in this paper has major limitations, the Bose-Fermi mapping method can only be applied to bosonic systems in the strong interaction limit, and its mathematical form can no longer satisfy the equations for the case of finite strong interactions. For the latter case we envision the partial use of the Bose-Fermi mapping method to construct orthogonal complete state spaces, so that the dynamical processes can continue to be obtained along the lines used in this paper. Finally, our work fills a small gap in the non-equilibrium dynamics of TG gases, and hope that the occupation distributions we obtain for different external drives will lead to new ideas for quantum state preparation. \\\\

\noindent{\bf Data availability}\\
The authors confirm that the data supporting the findings of this study are available within the article [and/or its supplementary materials].

\appendix
\section{Calculation of transition matrix elements}

In this appendix, we will give a detailed derivation of the transition matrix elements $\Gamma_{mn}$ in section \uppercase\expandafter{\romannumeral2}. Let's first consider one of the integral terms,
\begin{equation*}
\begin{aligned}
    \Gamma _{mn}^j =& \int {{\psi _m\left( {{x_1}, \cdots ,{x_N}} \right)}{x_j}{\psi _n\left( {{x_1}, \cdots ,{x_N}} \right)}} \prod\limits_{i = 1}^N {{\rm{d}}{x_i}} \\
    =& \left\langle {{m_1}{m_2} \cdots {m_{N}}} \right|{x_j}\left| {{n_1}{n_2} \cdots {n_{N}}} \right\rangle \\
    =& \frac{{{\alpha _0}}}{{\sqrt 2 }} \left\langle {{m_1}{m_2} \cdots {m_{N}}} \right|\left( {a_j^ +  + a_j^ - } \right)\left| {{n_1}{n_2} \cdots {n_{N}}} \right\rangle,
\end{aligned}
\end{equation*}
\begin{equation*}
    \left| {{n_1}{n_2} \cdots {n_N}} \right\rangle  = \frac{1}{{\sqrt {N!} }}\left| {\begin{array}{*{20}{c}}
    {{{\left| {{n_1}} \right\rangle }_1}}&{{{\left| {{n_1}} \right\rangle }_2}}& \cdots &{{{\left| {{n_1}} \right\rangle }_N}}\\
    {{{\left| {{n_2}} \right\rangle }_1}}&{{{\left| {{n_2}} \right\rangle }_2}}& \cdots &{{{\left| {{n_2}} \right\rangle }_N}}\\
     \vdots & \vdots & \ddots & \vdots \\
    {{{\left| {{n_N}} \right\rangle }_1}}&{{{\left| {{n_N}} \right\rangle }_2}}& \cdots &{{{\left| {{n_N}} \right\rangle }_N}}
    \end{array}} \right|,
\end{equation*}
where subscripts inside Dirac symbols represent the ordering of quantum states, while subscripts outside Dirac symbols represent the ordering of particles. The result of $x_i$ acting on a $N$ particles state $\left| {{n_1} \cdots {n_i} \cdots {n_{N}}} \right\rangle$ is
\begin{equation*}
\begin{aligned}
    &\frac{{{\alpha _0}}}{{\sqrt {2N!} }}\left| {\begin{array}{*{20}{c}}
    {{{\left| {{n_1}} \right\rangle }_1}}&{\sqrt {{n_1} + 1} {{\left| {{n_1} + 1} \right\rangle }_i}}&{{{\left| {{n_1}} \right\rangle }_N}}\\
    {{{\left| {{n_i}} \right\rangle }_1}}&{\sqrt {{n_i} + 1} {{\left| {{n_i} + 1} \right\rangle }_i}}&{{{\left| {{n_i}} \right\rangle }_N}}\\
    {{{\left| {{n_N}} \right\rangle }_1}}&{\sqrt {{n_N} + 1} {{\left| {{n_N} + 1} \right\rangle }_i}}&{{{\left| {{n_N}} \right\rangle }_N}}
    \end{array}} \right| \\
    +& \frac{{{\alpha _0}}}{{\sqrt {2N!} }}\left| {\begin{array}{*{20}{c}}
    {{{\left| {{n_1}} \right\rangle }_1}}&{\sqrt {{n_1} + 0} {{\left| {{n_1} - 1} \right\rangle }_i}}&{{{\left| {{n_1}} \right\rangle }_N}}\\
    {{{\left| {{n_i}} \right\rangle }_1}}&{\sqrt {{n_i} + 0} {{\left| {{n_i} - 1} \right\rangle }_i}}&{{{\left| {{n_i}} \right\rangle }_N}}\\
    {{{\left| {{n_N}} \right\rangle }_1}}&{\sqrt {{n_N} + 0} {{\left| {{n_N} - 1} \right\rangle }_i}}&{{{\left| {{n_N}} \right\rangle }_N}}
    \end{array}} \right|.
\end{aligned}
\end{equation*}

When taking the inner product of ${\left| {{m_1} \cdots {m_i} \cdots {m_N}} \right\rangle ^\dag }$ and $x_i\left| {{n_1} \cdots {n_i} \cdots {n_{N}}} \right\rangle$, expand the determinant of the first term above. This will result in $(N-1)!$ subterms with coefficients $\sqrt {{n_i} + 1}$. Let's focus only on the subterms with coefficients $\sqrt {{n_1} + 1}$,
\begin{equation*}
    \gamma_1 = \frac{{{\alpha _0}}}{{\sqrt {2N!} }}\sqrt {{n_1} + 1} {\left| {{n_1} + 1} \right\rangle _i}{\left| {{n_2} \cdots {n_N}} \right\rangle _i},
\end{equation*}
\begin{equation*}
    {\left| {{n_2} \cdots {n_N}} \right\rangle _i} = \left| {\begin{array}{*{20}{c}}
    {{{\left| {{n_2}} \right\rangle }_1}}&{{{\left| {{n_2}} \right\rangle }_{i - 1}}}&{{{\left| {{n_2}} \right\rangle }_{i + 1}}}&{{{\left| {{n_2}} \right\rangle }_N}}\\
    {{{\left| {{n_{i - 1}}} \right\rangle }_1}}&{{{\left| {{n_{i - 1}}} \right\rangle }_{i - 1}}}&{{{\left| {{n_{i - 1}}} \right\rangle }_{i + 1}}}&{{{\left| {{n_{i - 1}}} \right\rangle }_N}}\\
    {{{\left| {{n_i}} \right\rangle }_1}}&{{{\left| {{n_i}} \right\rangle }_{i - 1}}}&{{{\left| {{n_i}} \right\rangle }_{i + 1}}}&{{{\left| {{n_i}} \right\rangle }_N}}\\
    {{{\left| {{n_N}} \right\rangle }_1}}&{{{\left| {{n_N}} \right\rangle }_{i - 1}}}&{{{\left| {{n_N}} \right\rangle }_{i + 1}}}&{{{\left| {{n_N}} \right\rangle }_N}}
    \end{array}} \right|.
\end{equation*}

The inner product of ${\left| {{m_1} \cdots {m_i} \cdots {m_N}} \right\rangle ^\dag }$ and $\gamma_1$ is
\begin{equation*}
    \frac{{{\alpha _0}}}{{\sqrt 2 N}}\sqrt {{n_1} + 1} \left| {\begin{array}{*{20}{c}}
    {{\delta _{{n_1} + 1,{m_1}}}}&{{\delta _{{n_i},{m_1}}}}&{{\delta _{{n_N},{m_1}}}}\\
    {{\delta _{{n_1} + 1,{m_i}}}}&{{\delta _{{n_i},{m_i}}}}&{{\delta _{{n_N},{m_i}}}}\\
    {{\delta _{{n_1} + 1,{m_N}}}}&{{\delta _{{n_i},{m_N}}}}&{{\delta _{{n_N},{m_N}}}}
    \end{array}} \right|
\end{equation*}
and after adding up all the different coefficient terms we have
\begin{equation*}
\begin{aligned}
    & \left\langle {{m_1}{m_2} \cdots {m_{N}}} \right|{x_i}\left| {{n_1}{n_2} \cdots {n_{N}}} \right\rangle \\
    =& \frac{{{\alpha _0}}}{{\sqrt 2 }N}\sum\limits_{i = 1}^{N} {\sqrt {{n_i} + 1} \left| {\begin{array}{*{20}{c}}
    {{\delta _{{n_1},{m_1}}}}&{{\delta _{{n_i} + 1,{m_1}}}}&{{\delta _{{n_{N}},{m_1}}}}\\
    {{\delta _{{n_1},{m_i}}}}&{{\delta _{{n_i} + 1,{m_i}}}}&{{\delta _{{n_{N}},{m_i}}}}\\
    {{\delta _{{n_1},{m_{N}}}}}&{{\delta _{{n_i} + 1,{m_{N}}}}}&{{\delta _{{n_{N}},{m_{N}}}}}
    \end{array}} \right|} \\
    +& \frac{{{\alpha _0}}}{{\sqrt 2 }N}\sum\limits_{i = 1}^{N} {\sqrt {{n_i} + 0} \left| {\begin{array}{*{20}{c}}
    {{\delta _{{n_1},{m_1}}}}&{{\delta _{{n_i} - 1,{m_1}}}}&{{\delta _{{n_{N}},{m_1}}}}\\
    {{\delta _{{n_1},{m_i}}}}&{{\delta _{{n_i} - 1,{m_i}}}}&{{\delta _{{n_{N}},{m_i}}}}\\
    {{\delta _{{n_1},{m_{N}}}}}&{{\delta _{{n_i} - 1,{m_{N}}}}}&{{\delta _{{n_{N}},{m_{N}}}}}
    \end{array}} \right|}.
\end{aligned}
\end{equation*}
Only the calculation of $x_i$ is given and the result in section \uppercase\expandafter{\romannumeral 2} can be obtained by summing over all $x_i$.

\section{Density distribution and ROBDM}
The original expression of Eq. (\ref{eq4}) is
\begin{equation*}
\begin{aligned}
    D_{mn} =& \int \prod\limits_{i = 2}^N {{\rm{d}}{x_i}}{{\psi _m}\left( {x, \cdots ,{x_N}} \right){\psi _n}\left( {x, \cdots ,{x_N}} \right) } \\
    =& \int \prod\limits_{i = 2}^N {{\rm{d}}{x_i}} {\frac{1}{{N!}}} \left( {\sum\limits_P {{\varepsilon _P}{\varphi _{{p_1}}}\left( x \right) \cdots {\varphi _{{p_N}}}\left( {{x_N}} \right)} } \right) \\
    & \times \qquad\left( {\sum\limits_Q {{\varepsilon _Q}{\varphi _{{q_1}}}\left( x \right) \cdots {\varphi _{{q_N}}}\left( {{x_N}} \right)} } \right), \\
\end{aligned}
\end{equation*}
where ${\varepsilon _P}$ (${\varepsilon _Q}$) is the signature of $P=\left({{p_1}, \cdots ,{p_N}} \right)$ (Q=$\left( {{q_1}, \cdots ,{q_N}} \right)$) which corresponds to the permutation of the occupation configuration $\left( {{n_1},{n_2}, \cdots ,{n_N}} \right)$ ( $\left( {{m_1},{m_2}, \cdots ,{m_N}} \right)$) for $\left|\psi_m \right\rangle$ ($\left|\psi_n \right\rangle$). 
Due to the orthogonality of the single particle wave function $\varphi_l(x)$, the integral will be not zero only in two cases. For the case of $\left|{\psi _m}\right\rangle=\left|{\psi _n}\right\rangle$, the summation term is nonzero only for the term of $P=Q$. In other cases, $N-1$ particles must occupy in the same single particle levels for $\left|{\psi _m}\right\rangle$ and $\left|{\psi _n}\right\rangle$, i.e., only one element in the set of ${n_i}$ is not same as those in the set of ${m_j}$.

Finally, we give a simple calculation of equation (\ref{eq8}) as following
\begin{equation*}
\begin{aligned}
    P_{mn} =& \int {{\psi _m}\left( {x, \cdots ,{x_N}} \right){\psi _n}\left( {y, \cdots ,{x_N}} \right)} \\
    & \prod\limits_{j = 2}^N {{\rm{sign}}\left( {x - {x_j}} \right){\rm{sign}}\left( {y - {x_j}} \right)} \prod\limits_{i = 2}^N {{\rm{d}}{x_i}} \\
    =& \int \prod\limits_{j = 2}^{N - 1} {{\rm{sign}}\left( {x - {x_j}} \right) {\rm{sign}}\left( {y - {x_j}} \right)} {\prod\limits_{i = 2}^{N - 1} {{\rm{d}}{x_i}} } \\
    & \quad\int_{ - \infty }^{ + \infty } {{\psi _m}\left( {x, \cdots ,{x_N}} \right){\psi _n}\left( {y, \cdots ,{x_N}} \right)}{\rm{d}}{x_N} \\
    -& 2\int \prod\limits_{j = 2}^{N - 1} {{\rm{sign}}\left( {x - {x_j}} \right) {\rm{sign}}\left( {y - {x_j}} \right)} {\prod\limits_{i = 2}^{N - 1} {{\rm{d}}{x_i}} } \\
    & \quad\int_x^y {{\psi _m}\left( {x, \cdots ,{x_N}} \right){\psi _n}\left( {y, \cdots ,{x_N}} \right)}{\rm{d}}{x_N},
\end{aligned}
\end{equation*}
where the signature on $x_N$ has been removed. Repeating the above process we can have the formula in the main text. To show the process of the above calculation more clearly, we give the example $P_{12}$ for two particles system, i.e., $\left|\psi _1\right\rangle = \left| {0,1} \right\rangle$ and $\left|\psi _2\right\rangle = \left| {0,2} \right\rangle$
\begin{equation*}
\begin{aligned}
    \int_{ - \infty }^{ + \infty } & {{\psi _1}\left( {x,{x_2}} \right)} {\psi _2}\left( {y,{x_2}} \right){\rm{sign}}\left( {x - {x_2}} \right){\rm{sign}}\left( {y - {x_2}} \right){\rm{d}}{x_2} \\
    & = \int_{ - \infty }^{ + \infty } {{\psi _1}\left( {x,{x_2}} \right)} {\psi _2}\left( {y,{x_2}} \right){\rm{d}}{x_2} \\
    & \quad - 2\int_x^y {{\psi _1}\left( {x,{x_2}} \right)} {\psi _2}\left( {y,{x_2}} \right){\rm{d}}{x_2} \\
    & = {\varphi _0}\left( x \right){\varphi _0}\left( y \right)\left[ {0 - 2\int_x^y {{\varphi _1}\left( {{x_2}} \right)} {\varphi _2}\left( {{x_2}} \right){\rm{d}}{x_2}} \right] \\
    & - {\varphi _0}\left( x \right){\varphi _2}\left( y \right)\left[ {0 - 2\int_x^y {{\varphi _1}\left( {{x_2}} \right)} {\varphi _0}\left( {{x_2}} \right){\rm{d}}{x_2}} \right] \\
    & - {\varphi _1}\left( x \right){\varphi _0}\left( y \right)\left[ {0 - 2\int_x^y {{\varphi _0}\left( {{x_2}} \right)} {\varphi _2}\left( {{x_2}} \right){\rm{d}}{x_2}} \right] \\
    & + {\varphi _1}\left( x \right){\varphi _2}\left( y \right)\left[ {1 - 2\int_x^y {{\varphi _0}\left( {{x_2}} \right)} {\varphi _0}\left( {{x_2}} \right){\rm{d}}{x_2}} \right].
\end{aligned}
\end{equation*}

\bibliography{JPBResub}

\end{document}